\documentclass[pre,nofootinbib,floatfix,superscriptaddress, twocolumn]{revtex4}
\usepackage{dcolumn}
\usepackage{graphicx}
\usepackage{amsmath,amsfonts,amssymb, epsfig}
\usepackage{amsthm}
\usepackage{natbib}
\usepackage[outdir=../used_figures/]{epstopdf}
\usepackage{verbatim}
\usepackage[caption=false, position=top]{subfig}
\usepackage{color}
\usepackage{rotating}
\usepackage{float}
\usepackage{subfiles}
\usepackage{comment}
\graphicspath{{figures/}{/../figures/}}

\newcommand{\msum}{\sum_{m=1}^N}
\newcommand{\pvec}{\mathbf{p}}

\newcommand{\bvec}{\mathbf{b}}

\newcommand{\cvec}{\mathbf{c}}
\newcommand{\muvec}{\mathbf{m}}
\newcommand{\onevec}{\mathbf{1}}
\newcommand{\inv}{^{-1}}

\newcommand{\hats}{\hat{S}}

\newcommand{\kmin}{k_{\text{min}}}
\newcommand{\kmax}{k_{\text{max}}}
\newcommand{\kinu}{\mathbf{k}^{\text{in}}}
\newcommand{\kinl}{\mathbf{k}_{\text{in}}}
\newcommand{\koutu}{\mathbf{k}^{\text{out}}}
\newcommand{\koutl}{\mathbf{k}_{\text{out}}}

\usepackage{xargs}                      
\usepackage[pdftex,dvipsnames]{xcolor}  
\usepackage[colorinlistoftodos,prependcaption,textsize=tiny]{todonotes}

\usepackage[pdftex,bookmarks, colorlinks, citecolor =blue, breaklinks]{hyperref}

\begin{document}
\title{Control of excitable systems is optimal near criticality}
\author{Kathleen Finlinson}
	\affiliation{Department of Applied Mathematics, University of Colorado at Boulder, Boulder, CO, 80309, USA}
\author{Woodrow L. Shew}
	\affiliation{Department of Physics, University of Arkansas, Fayetteville, Arkansas, 72701, USA}
\author{Daniel B. Larremore}
	\affiliation{Department of Computer Science, University of Colorado at Boulder, Boulder, CO, 80309, USA}
	\affiliation{BioFrontiers Institute, University of Colorado at Boulder, Boulder, CO, 80303, USA}
\author{Juan G. Restrepo}
	\email{juanga@colorado.edu}
	\affiliation{Department of Applied Mathematics, University of Colorado at Boulder, Boulder, CO, 80309, USA}

\begin{abstract}
Experiments suggest that cerebral cortex gains several functional advantages by operating in a dynamical regime near the critical point of a phase transition.  However, a long-standing criticism of this hypothesis is that critical dynamics are rather noisy, which might be detrimental to aspects of brain function that require precision.  If the cortex does operate near criticality, how might it mitigate the noisy fluctuations?  One possibility is that other parts of the brain may act to control the fluctuations and reduce cortical noise.  To better understand this possibility, here we numerically and analytically study a network of binary neurons.  We determine how efficacy of controlling the population firing rate depends on proximity to criticality as well as different structural properties of the network.  We found that control is most effective - errors are minimal for the widest range of target firing rates - near criticality.  Optimal control is slightly away from criticality for networks with heterogeneous degree distributions.  Thus, while criticality is the noisiest dynamical regime, it is also the regime that is easiest to control, which may offer a way to mitigate the noise. 
\end{abstract}

\maketitle

A large body of experimental evidence supports the hypothesis that the cerebral cortex operates at, or very near to, a critical regime in which excitation and inhibition are precisely balanced \cite{yu2017maintained, scott2014voltage, shew2015adaptation, bellay2015irregular, ma2019cortical}. Consequently, many efforts have been made to identify and understand the functional benefits of operating in this regime \cite{shew2013functional}, including maximized dynamic range \cite{kinouchi2006optimal, shew2009neuronal, gautam2015maximizing} and maximized information transmission and capacity \cite{shew2011information, fagerholm2016cortical, clawson2017adaptation}. Both experiments and models have shown that networks operating  in the critical regime visit the largest variety of microscopic and macroscopic states \cite{shew2011information,  larremore2014inhibition, agrawal2018robust}. It has been hypothesized that these states could be harnessed by the network to encode and transfer information \cite{luczak2009spontaneous}. 
In this Letter, we identify a new way that criticality may be beneficial in neural systems.  We show that, at criticality, the activity of the system can be controlled with minimal error over the largest range of activity levels.
In addition, by analytically treating the network's deviations from linear dynamics, we show that heterogeneity in the network's degree distribution reduces this controllable range considerably. Understanding the controllability of a neural system may be important for designing effective neuroprostetics and other brain-machine interface systems. There has been a significant amount of work in quantifying the controllability of functional brain networks (for a review, see \cite{lynn2019physics}). However, these previous studies assume linear dynamics and consider a fixed excitation-inhibition balance. In contrast, in this Letter we consider the combined effects of nonlinear dynamical evolution, network heterogeneity, and excitation-inhibition balance. 

Following Refs.~\cite{kinouchi2006optimal,larremore2014inhibition, agrawal2018robust}, we consider a network of $N$ discrete-state excitable nodes. At each timestep $t$, each node $m$ may be active or inactive, corresponding to a node variable $s_m^t$ with value $1$ or $0$, respectively. The nodes are coupled via a directed, weighted, and signed network with adjacency matrix $A$, such that $A_{nm}$ represents the stimulus that an active node $m$ sends to node $n$. These values are non-negative (non-positive) whenever node $m$ is excitatory (inhibitory), and networks are assumed to consist of 20\% inhibitory nodes. In a system with no control, dynamics then autonomously evolve with node $n$ becoming active at timestep $t+1$ with probability $\sigma\left(\msum A_{nm}s_m^t\right)$, where $\sigma$ is a piecewise linear transfer function defined as $\sigma(x)=0$ for $x<0$, $\sigma(x)=x$ for $0 \leq x \leq 1$, and $\sigma(x)=1$ for $x>1$. In such systems, the largest eigenvalue $\lambda$ of the adjacency matrix $A$ determines whether the system is in a regime of low activity ($\lambda < 1$), high activity ($\lambda > 1$), or a state separating those regimes ($\lambda=1$) characterized by high entropy of the macroscopic network activity $S^t = \frac{1}{N} \sum_{m=1}^N s_m^t$~\cite{larremore2014inhibition, agrawal2018robust}. 
 
We now introduce control to the dynamics and analyze how its effectiveness depends on the network's topology and principal eigenvalue $\lambda$. To this end, we consider a modified evolution equation, 
\begin{equation}
	s_n^{t+1}\!=\!1\text{ w.p. }\sigma\bigg[\sum_{m=1}^N A_{nm}s_m^t\!+\!\mu_n\bigg(\hats\!-\!\sum_{m=1}^Nb_ms_m^t\bigg)\bigg]
	\label{eq:global_control_evolution}
\end{equation}
and $s_n^{t+1}=0$ otherwise. 
The term $ \mu_n(\hats-\sum_{m=1}^Nb_ms_m^t)$ represents an external proportional control signal which attempts to bring the global variable $S = \msum b_m s_m^t$ to a target value $\hats$, where $b_n$ represents the weight given to node $n$ in the global activity $S$. Purposefully, this control goal is simpler than those considered in some previous works \cite{lynn2019physics}; this simplicity allows us to focus on the effects of the nonlinearity, the network's degree distribution, and excitation-inhibition balance. As a first step, we will ignore the nonlinearities of the function $\sigma$ by assuming that its argument is always between $0$ and $1$, the regime in which $\sigma(x) = x$. Later we will discuss the conditions under which this assumption is valid and consider the effects of the nonlinearity.

In this linear regime, the evolution of each node's expected activity $p_n^t \equiv E[s_n^t]$, taken over the stochastic dynamics, is, in steady-state,
\begin{equation}
	\pvec = A \pvec + \muvec \left(\hats  - \bvec^\top \pvec \right)\ . 
	\label{eq:linear}
\end{equation}
where we have introduced vectors ${\mathbf m}$, $\bvec$, and $\pvec$ with entries $\mu_n$, $b_n$, and $p_n$, respectively.
This can be solved for $\pvec$ using the Sherman-Morrison inversion formula to get
\begin{equation}
	\pvec = \left( (I-A)\inv - \frac{(I-A)\inv \muvec \bvec^\top (I-A)\inv}{1+ \bvec^\top (I-A)\inv \muvec} \right) \hats \muvec\ . \nonumber
\end{equation}
Defining $\cvec = (I-A)\inv \muvec$ simplifies this equation to
\begin{align}
  \pvec &= \hats \cvec \left( 1- \frac{\bvec^\top \cvec}{1+\bvec^\top \cvec}  \right)  
     = \frac{\hats \cvec}{ 1+ \bvec^\top \cvec}\ . \label{eq:p solution}
\end{align}
Equation~\eqref{eq:p solution} predicts the steady-state averages of nodal activity $p_n = E[s_n^t]$ in terms $\cvec$, and therefore in terms of the network $A$ and the control vector $\muvec$. 

How does control error depend on the network's degree distribution and proximity to the critical point $\lambda\!=\!1$? We first assume that the in- and out-degrees, $k^{\text{out}}_n = \msum A_{mn}$ and $k^{\text{in}}_n = \msum A_{nm}$, and the largest eigenvalue, $\lambda$, are specified. Then, following \cite{sonnenschein2012onset}, we consider an averaged adjacency matrix $\bar A$ which preserves the eigenvalue $\lambda$ and the degree distribution, in the sense that $\msum \bar A_{nm} \propto k^{\text{in}}_n$, $\msum \bar A_{mn} \propto k^{\text{out}}_n$. Such a matrix is $\bar A = \lambda \kinl \koutl ^\top/\koutl^\top \kinl$, which has the eigenvector $\kinl$ with eigenvalue $\lambda$. Using this matrix, we effectively average over the ensemble of networks with largest eigenvalue $\lambda$ and the desired degree distributions.
\begin{figure}[t]
	\centering
	\includegraphics[width=1.0\columnwidth]{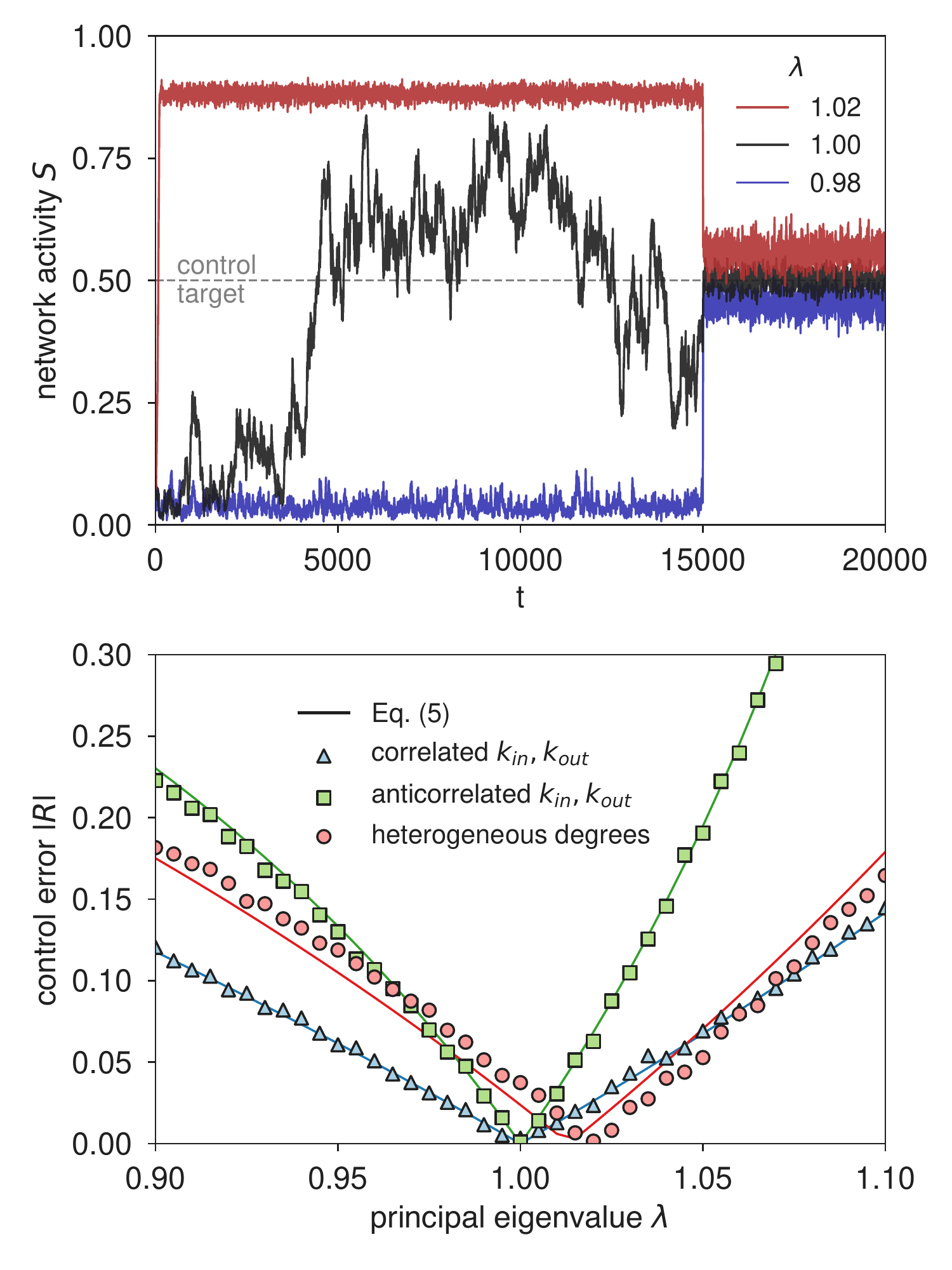}
	\caption{(A) Network activity $S^t$ as a function of $t$ for three values of $\lambda$ as indicated, with and without control. Control is turned on at $t = 15,000$, with target $\hat S = 0.5$ (dashed line), using an Erd\"os-R\'enyi network of size $N = 5000$ and mean degree $200$ with $\muvec = \tfrac{1}{2} \onevec$ and $\bvec = \tfrac{1}{N}\onevec$. (B) Absolute control error obtained numerically from direct evolution of Eqs.~\eqref{eq:global_control_evolution} (symbols) and theoretically from Eq.~\eqref{eq:homogeneous error} (solid lines) for networks with different correlations (see text).}
 \label{fig:1}
\end{figure}
Replacing $A$ with $\bar A$ in the definition of $\cvec$, and using the Sherman-Morrison inversion formula again, we get
\begin{align}
	\cvec &= \left( I + \frac{\lambda \kinl \koutl^\top}{(1-\lambda) \koutl^\top \kinl} \right) \muvec \ .
	\label{eq:c definition}
\end{align}
We quantify control error as the expected relative error  ${R = (\bvec^\top \pvec - \hats)/\hat S}$. Substituting Eq.~\eqref{eq:p solution} for $\pvec$ we find ${R = -1/(1 + \bvec^\top\cvec)}$. Then, inserting Eq.~\eqref{eq:c definition} for $\cvec$ yields 
\begin{align}
	R(\lambda) &= \frac{ (\lambda -1) }
                         {
                         (1-\lambda)(1+\bvec^\top \muvec) +
                         \lambda\frac{\bvec^\top \kinl \koutl^\top \muvec}{\koutl^\top \kinl}
                        }\ .
	\label{eq:homogeneous error}
\end{align}

This equation makes two key predictions for the control error $R$, based on the network's degree distribution $(\kinu,\koutu)$, the control vector $\muvec$, the nodal weights $\bvec$, and the principal eigenvalue $\lambda$.  First, Eq.~\eqref{eq:homogeneous error} predicts that control error is minimized to $R=0$ whenever $\lambda=1$, independent of the target $\hat{S}$. Second, Eq.~\eqref{eq:homogeneous error} predicts that when $\lambda \neq 1$, the magnitude of the non-zero control error depends on correlations between the in- and out-degrees of the nodes, the weights $\bvec$, and the nodal control strengths $\muvec$. We confirm these predictions via simulation by varying both the principal eigenvalue of the network adjacency matrix and the correlation between in- and out-degree sequences. Figure~\ref{fig:1}A illustrates the effects of $\lambda$ on dynamics with and without control, with sample time series of network activity $S^t = \sum_{n=1}^N s_n^t/N$ for simulated dynamics with $\lambda = 0.98$, $1.0$, and $1.02$ on directed Erd\"os-R\'enyi random networks (simulation details in caption). Figure.~\ref{fig:1}B shows the accuracy of Eq.~\eqref{eq:homogeneous error} (solid lines) compared to simulations (squares and triangles) in which we systematically varied $\lambda$ between 0.9 and 1.1 for two classes of networks which were designed to maximize and minimize the effects of degree correlations, respectively. To construct such networks, we considered a sequence of target degrees $\hat k_1 > \hat k_2 > \dots > \hat k_N$ with $N = 2000$ sampled from a distribution uniform in $[50, 250]$. To maximize the term $\bvec^\top \kinl \koutl^\top \muvec/\koutl^\top \kinl$, we chose $k^{out}_n = \hat k_n$, $k^{in}_n = \hat k_{N-n}$, $\bvec \propto \kinu$, and $\muvec \propto \koutu$; to minimize that term, we chose $k^{out}_n = k^{in}_n = \hat k_n$ and $\mu_n, b_n \propto \hat k_{N-n}$. We generated adjacency matrices $A$ by choosing the entries $A_{nm}$ to be $1$ with probability $k^{in}_n k^{out}_m/ \sum_{n} k^{out}_n$ \cite{chunglu}, and then rescaled the resulting matrices so that they had the desired eigenvalue $\lambda$. We emphasize that while the simulations in Fig.~\ref{fig:1}B were done with specific network realizations, the predictions of Eq.~\eqref{eq:homogeneous error} were derived by considering an averaged network.

\begin{figure}[t]
  \includegraphics[width=\columnwidth]{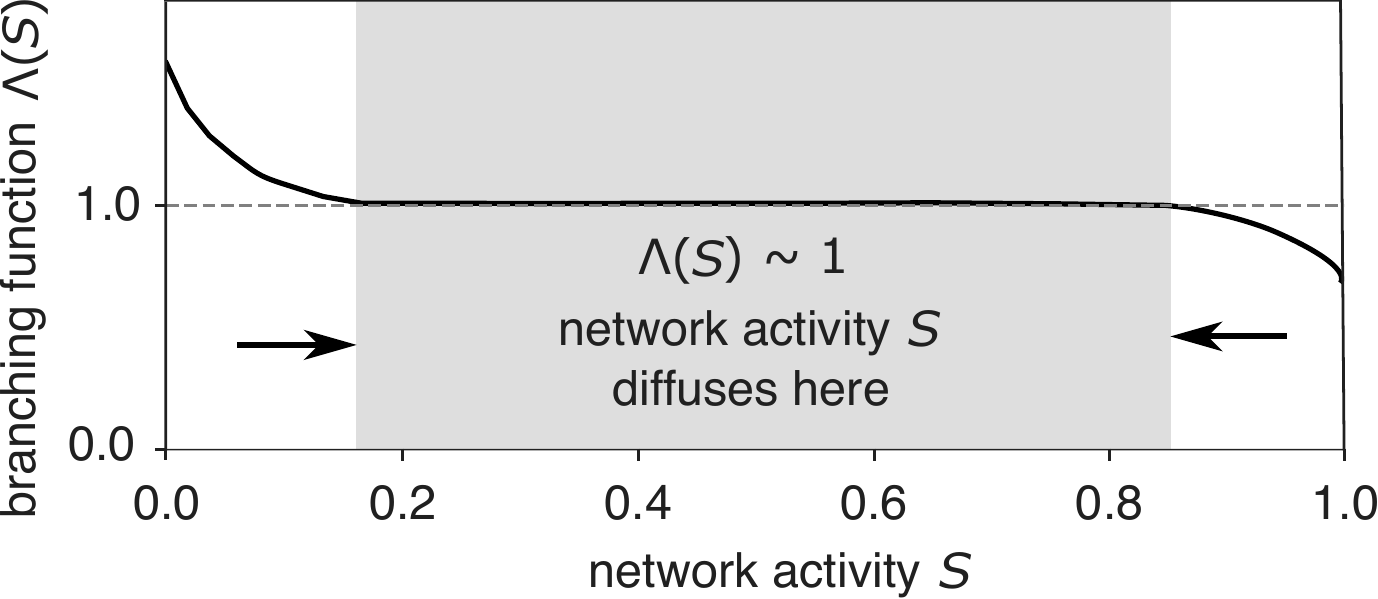}
    \caption{Schematic representation of the branching function $\Lambda(S) = E[S^{t+1}|S^t = S]/S$. Activity tends to increase [decrease] on average when $\Lambda(S) > 1$ [$\Lambda(S) < 1$].}
    \label{fig:cartoon}
\end{figure}

In the two cases described above, the control error is well described by Eq.~(\ref{eq:homogeneous error}), but this is not the case if the network's degree distribution is heterogeneous (even mildly). For example, if the degrees $\hat k_n$ are sampled from a power-law distribution with minimum degree $50$ and exponent $\gamma = 4$ (circles in Fig.~\ref{fig:1}B), the control error is no longer minimized at $\lambda = 1$. As we will see, the assumption that the argument $x$ of the sigmoid function $\sigma(x)$ is between $0$ and $1$ is violated when the network's degree distribution is heterogeneous. This assumption allowed us to ignore the nonlinearity of $\sigma$ and write down the linear equation Eq.~\eqref{eq:linear}. To consider the effects of nonlinearity we will assume for simplicity that each node is weighted equally, i.e., $b_n = 1/N$, and consider the evolution of the network activity $S^t = \sum_{n=1}^N s_n^t/N$. In the absence of control input and nonlinearity, the expected network activity for a homogeneous network with largest eigenvalue $\lambda$ evolves, given $S^t$, as $E[S^{t+1}] = \lambda S^t$; the largest eigenvalue $\lambda$ acts as a branching parameter for the fraction of excited nodes. Thus, our linear results above can be understood as follows: when $\lambda = 1$, $S^t$ diffuses stochastically with no bias, and therefore control can effectively pin it to any value of the target $\hat S$. When the nonlinearity of the transfer function is taken into account one can still quantify the bias of the stochastic evolution of $S^t$ by defining the {\it branching function} $\Lambda$ that, given $S^t$, satisfies $E[S^{t+1}] = \Lambda S^t$:
\begin{equation}
	\Lambda(S) = E\left [S^{t+1}\mid S^t=S\right]/S.
\end{equation}
The branching function $\Lambda(S)$ can be interpreted as an effective eigenvalue or branching ratio that applies when $S^t = S$: when $\Lambda(S) > 1$ [$\Lambda(S) < 1$] activity tends to increase [decrease] on average. Alternatively, interpreting the evolution of $S^t$ as a random walk, its drift is given by $E[S^{t+1}-S^t \mid S^t\! =\! S] = \Lambda(S)(S-1)$. Therefore, when the branching function $\Lambda(S) \approx 1$, the network activity $S^t$ does not have a tendency to increase or decrease. In Ref.~\cite{larremore2014inhibition} it was shown that for directed Erd\"os-R\'enyi networks, depending on the fraction of inhibitory neurons and the average network degree, the branching function is a decreasing function of $S$ with $\Lambda(S) \geq \lambda$ for small $S$, $\Lambda(S) \approx \lambda$ for intermediate $S$, and $\Lambda(S) < \lambda$ for $S \approx 1$. For networks with $\lambda = 1$, $S^t$ effectively diffuses in the region where $\Lambda(S) \approx 1$ (Fig.~\ref{fig:cartoon}). If the target $\hats$ is in this region, $S^t$ can be controlled to be at this value with effectively no error.  Our goal now is to study how a heterogeneous degree distribution modifies the branching function, and consequently the range of values $\hats$ that can be used as targets with no error. 


From the definition of the branching function and Eq.~\eqref{eq:global_control_evolution}, we obtain
\begin{align}
  \Lambda(S)  = \frac{1}{S N} E\left[\sum_{n=1}^N \sigma\left(\sum_{m=1}^N A_{nm} s_m^t\right) \big|S^t=S\right]
\end{align}
If the number of connections per node is large, then the distribution of $\sum_{m=1}^N A_{nm}s_m^t$ is sharply peaked around its mean, $\sum_{m=1}^N \bar A_{nm}  S = k^{in}_n \lambda S/ 
\langle k \rangle$, where $\langle k \rangle \equiv \sum_{n=1}^N k^{out}_n/N$, and assuming that $s_m^t$ is uncorrelated with $k^{out}_m$. Therefore we get $\Lambda(S) \approx  \sum_{n=1}^N \sigma\left(k^{in}_n \lambda S/ \langle k \rangle\right)/(NS)$. Expressing this sum as an integral over the in-degree distribution $P(k)$ and using $\sigma(x) = x$ for $x \leq 1$, $\sigma(x) = 1$ for $x> 1$, we obtain an expression for the branching function $\Lambda(S)$:
\begin{align}
  \Lambda(S) &= \frac{\lambda}{\langle k \rangle} \int_{\kmin}^{\langle k \rangle/(\lambda S)}
              k P(k) dk
  + \frac{1}{S} \int_{\langle k \rangle/(\lambda S)}^{\kmax} P(k) dk. \label{eq:Lambda} 
\end{align}
Note that if $\kmax \leq \langle k \rangle/(\lambda S)$, then Eq.~\eqref{eq:Lambda} collapses to $\Lambda(S) = \lambda$, showing how the nonlinear case reduces to the linear one, but if $\kmax > \langle k \rangle/(\lambda S)$, then $\Lambda(S) < \lambda$. This means that the more heterogeneous the in-degree distribution, the more $\Lambda(S)$ differs from $\lambda$. For the case $\lambda = 1$, $\Lambda(S) \approx 1$ as long as $S < \langle k \rangle/\kmax$. Therefore the range of target values $\hat S$ that can be controlled with zero error  is reduced for heterogeneous networks. When $\bvec = \tfrac{1}{N}\onevec$, the definition of error $R$ is equivalent to $S = \hat S + R(\lambda) \hat S$. Using $\Lambda(S)$ instead of $\lambda$ in Eq.~\eqref{eq:homogeneous error} gives a heuristic approximation for the value of $S$ as the solution of the implicit equation
\begin{align}\label{eq:nonli}
S = \hat S + R\left(\Lambda(S)\right) \hat S,
\end{align}
from which the relative error can be computed.

\begin{figure}[t]
	\centering
	\includegraphics[width=\columnwidth]{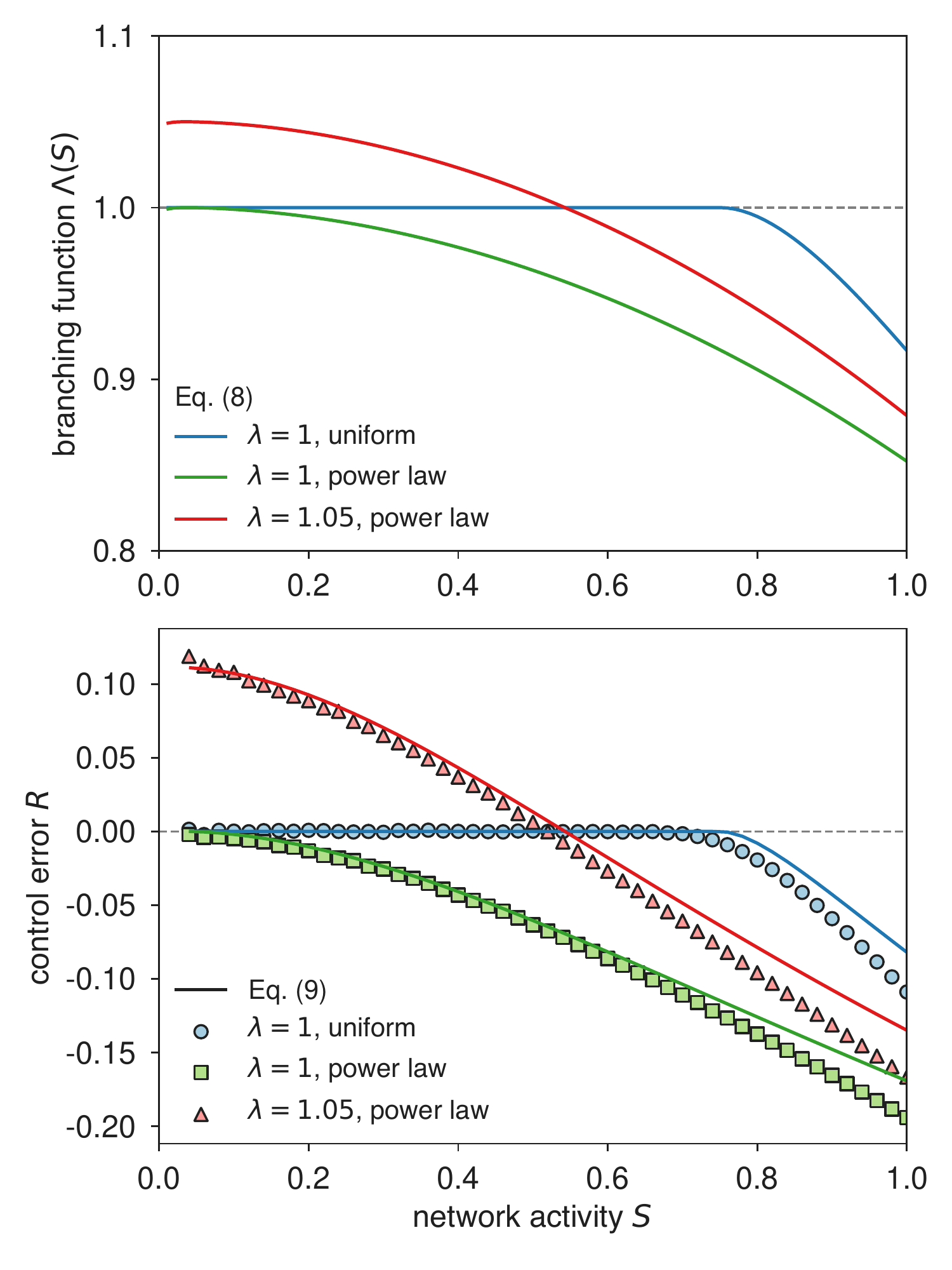}
	\caption{Experiments with three networks with $N\!=\!5000$ show validity of Eqs.~\eqref{eq:Lambda} and ~\eqref{eq:nonli}. Network degree distributions were either uniform in $[100,200]$ (blue, $\lambda=1$) or power laws with $\gamma=4$ and minimum degree $50$ (green, $\lambda=1$; red, $\lambda=1.05$).  (top) Branching functions calculated using Eq.~\eqref{eq:Lambda} for the three networks. (bottom) Control error measured numerically (symbols) and theoretically using $S$ obtained from Eq.~\eqref{eq:nonli} (lines) as a function of target $\hats$ for the same three networks. Note that error is small whenever the branching function is close to $1$.}
	\label{fig:2}
\end{figure}

We illustrate how heterogeneity in the degree distribution modifies the controllable range of targets by considering networks of size $N = 5000$ generated as before, but using $\bvec = \tfrac{1}{N}\onevec$, $\muvec = \tfrac{1}{2} \onevec$, $\{ \hat k_n \}$ sampled from either a power-law distribution with exponent $\gamma = 4$ with minimum degree $k_{\text{min}} = 50$ or a uniform distribution in $[100,200]$, and we choose $\{k^{in}_n\}$, $\{k^{out}_n\}$ as two independent permutations of the $\{\hat k_n\}$ sequence. (In light of recent discussion \cite{broido2019scale}, note that we choose power-law degree distributions just as a convenient example of heavy tailed distributions.) Fig.~\ref{fig:2}a shows branching functions $\Lambda(S)$ calculated using Eq.~\eqref{eq:Lambda} for three combinations of $\lambda$ and degree distributions (see caption). Of the two networks with $\lambda = 1$, the most homogeneous network has a branching function which is approximately $1$ for a wide range of values of $S$. For the network with $\lambda = 1.05$, the branching function crosses $1$ only at a value of $S$ approximately equal to $0.53$, effectively reducing the controllable range to a single point. All this is reflected in the error $R$, plotted as a function of the target $\hat S$ in Fig.~\ref{fig:2}b. In all cases, the region where the error is approximately zero corresponds to the region where the branching function is approximately one. In addition, the error computed theoretically from Eq.~\eqref{eq:nonli} (symbols) agrees qualitatively with the one obtained numerically from simulations of the full system (solid line). Networks with a homogeneous degree distribution and $\lambda = 1$ have the largest controllable range. We also find that networks with a more homogeneous degree distribution admit a larger range of control strengths $\mu$ before the controlled state becomes unstable. As shown in the Supplemental Material, we find that a condition for stability is 
\begin{align}
\mu < \frac{\lambda + 1}{ \lambda + 1 - h \lambda},\label{eq:munstable}
\end{align}
 where $h \equiv \langle k \rangle^2/\langle k^2 \rangle \leq 1$ measures the network's degree heterogeneity. More heterogeneous (smaller $h$) networks are therefore harder to control.

In summary, we studied how two key network properties (excitability and degree distribution heterogeneity) affect the ability to control the macroscopic activity of a network of excitable elements with inhibition. We found that for networks poised at the critical point of a phase transition ($\lambda = 1$), there is a relatively large range of macroscopic network activity states that can be stabilized with no error. Heterogeneity in the network's degree distribution reduces this range of target values and the range of control strengths that yield stable control. While heterogeneity can be beneficial for robustness to random node failures \cite{albert2000error}, our results suggest that a more homogeneous degree distribution might be preferable for situations where a large range of macroscopic network states needs to be harnessed.  A common critique of the hypothesis that the cerebral cortex may operate near criticality is that critical dynamics are too noisy, as reflected in the large fluctuations in Fig 1A.  For many aspects of brain function it is easy to imagine that the large fluctuations of criticality would cause trouble.  However, our primary result here is that the noisy dynamics of criticality is, in fact, easy to control.  This suggests that a brain might be able to take advantage of the other functional benefits of criticality while controlling its own noise to remain at a manageable level.

\bibliographystyle{plain}

\begin{thebibliography}{99}
\bibitem{yu2017maintained} S. Yu, T. L. Ribeiro, C. Meisel, S. Chou, A. Mitz, R. Saunders, and D. Plenz, ELife {\bf 6}, e27119 (2017).
\bibitem{scott2014voltage} G. Scott, E. D. Fagerholm, H. Mutoh, R. Leech, D. J. Sharp, W. L. Shew, and T. Kn\"opfel, Journal of Neuroscience {\bf 34}, 16611 (2014).
\bibitem{shew2015adaptation} W. L. Shew, W. P. Clawson, J. Pobst, Y. Karimipanah, N. C. Wright, and R. Wessel, Nature Physics  {\bf 11}, 659 (2015).
\bibitem{bellay2015irregular} T. Bellay, A. Klaus, S. Seshadri, and D. Plenz, ELife {\bf 4}, e07224 (2015).
\bibitem{ma2019cortical} Z. Ma, G. G. Turrigiano, R. Wessel, and K. B. Hengen, Neuron {\bf 104}, 655 (2019).
\bibitem{shew2013functional} W. L. Shew and D. Plenz, The Neuroscientist {\bf 19}, 88 (2013).
\bibitem{kinouchi2006optimal} O. Kinouchi and M. Copelli, Nature Physics {\bf 2}, 348 (2006).
\bibitem{shew2009neuronal} W. L .Shew, H. Yang, T. Petermann, R. Roy, and D. Plenz, Journal of Neuroscience {\bf 29}, 15595 (2009).
\bibitem{gautam2015maximizing} S. H. Gautam, T. T. Hoang, K. McClanahan, S. K. Grady, and W. L. Shew, PLoS Computational Biology {\bf 11}, e1004576 (2015).
\bibitem{shew2011information} W. L. Shew, H. Yang, S. Yu, R. Roy, and D. Plenz, Journal of Neuroscience {\bf 31}, 55 (2011).
\bibitem{fagerholm2016cortical} E. D. Fagerholm, G. Scott, W. L. Shew, C. Song, R. Leech, T. Kno\''pfel, and D. J. Sharp, Cerebral cortex {\bf 26} 3945 (2016).
\bibitem{clawson2017adaptation} W. P. Clawson, N. C. Wright, R. Wessel, and W. L. Shew,  PLoS computational biology {\bf 13}, e1005574 (2017).
\bibitem{larremore2014inhibition} D. B. Larremore, W. L. Shew, E. Ott, F. Sorrentino, and J. G. Restrepo, Physical Review Letters {\bf 112} 138103  (2014).
\bibitem{agrawal2018robust} V. Agrawal, A. B. Cowley, Q. Alfaori, D. B. Larremore, J. G Restrepo, and W. L. Shew, Chaos: An Interdisciplinary Journal of Nonlinear Science {\bf 28}, 103115  (2018).
\bibitem{luczak2009spontaneous} A. Luczak, P. Bartho\', and K. D. Harris, Neuron {\bf 62} 413 (2009).
\bibitem{lynn2019physics} C. W. Lynn and D. S. Bassett, Nature Reviews Physics {\bf 1}, 318 (2019).
\bibitem{sonnenschein2012onset} B. Sonnenschein and L. Schimansky-Geier, Physical Review E {\bf 85}, 051116 (2012).
\bibitem{chunglu} F. Chung and L. Lu, Annals of Combinatorics {\bf 6},125 (2002).
\bibitem{broido2019scale} A. D. Broido and A. Clauset, Nature Communications {\bf 10}, 1017 (2019).
\bibitem{albert2000error} R. Albert, H. Jeong, and A.-L. Barab\'asi, Nature {\bf 406} 378 (2000).

\end{thebibliography}

\end{document}